\documentclass{aa}  
\usepackage{multirow}
\usepackage{multicol}
\usepackage{bm}
\usepackage{imakeidx} 
\usepackage[T1]{fontenc}
\usepackage{siunitx} 
\usepackage{graphicx} 
\usepackage[table,xcdraw]{xcolor} 
\usepackage{caption}
\usepackage{subcaption}
\usepackage{amsmath} 
\usepackage{geometry}
\usepackage{float} 
\usepackage{here} 
\usepackage{xcolor} 
\usepackage{inputenc}
\usepackage[colorlinks=true, pdfstartview=FitV, linkcolor=blue, citecolor=blue, urlcolor=blue, breaklinks=true]{hyperref}

\usepackage{pdfpages}
\usepackage{hyphenat}

\def\Teff{T_{\rm eff}}

\def\Ro{R_{\odot}}
\def\deg{\ensuremath{^\circ}}

\newcommand{\Gaia}{{\textit{Gaia}}}
\newcommand{\gspspec}{{\textit{GSP-Spec}}}

\newcommand{\mh}{[M/H]}
\newcommand{\feh}{[Fe/H]}

\newcommand{\cafe}{[Ca/Fe]}

\newcommand{\ebprp}{E($B_{P}$ - $R_{P}$)}
\newcommand{\ebv}{E(B - V)}

\usepackage{color}

\usepackage{cleveref}
\usepackage{tabularx,booktabs}

\begin{document}


\title{\textbf{3D extinction maps of the Milky Way disc from \Gaia\ \gspspec\ parameters}}

\author{M. Barbillon \inst{1}, A. Recio-Blanco \inst{1}, P. de Laverny \inst{1}, P. A. Palicio\inst{1}}
\institute{\inst{1}\textit{Université Côte d’Azur, Observatoire de la Côte d’Azur, CNRS, Laboratoire Lagrange, France}}

\date{Received 14 November 2025; Accepted XXX}


\abstract{
Three-dimensional maps of interstellar dust are crucial for understanding the structure of the interstellar medium of the Milky Way and for applying correction to astrophysical observations affected by dust. Considerable progress has been achieved in this field thanks to the \Gaia\ mission.} 
{We aim at providing new extinction estimates in the $B_P$ and $R_P$ bands of the \Gaia\ mission to study the dust distribution in the Milky Way disc, to provide new views of the spatial distribution of interstellar extinction and to compare it with different tracers of the Galactic spiral arms.}
{We use a highly homogeneous method based on the spectral chemo-physical parametrisation of stars from \Gaia\ General Stellar Parametriser from Spectroscopy (\gspspec). This catalogue, composed of 5.6 million sources in the \Gaia\ Data Release 3, presents the advantage of estimating the stellar atmospheric parameters independently of extinction.
The \ebprp\ extinction is calculated by comparing the observed stellar $(B_P - R_P)$ colours in the \Gaia\ bands with the theoretical ones assuming a theoretical $\Teff$-log($g$)-\mh–colour relation, from the \gspspec\ stellar atmospheric parameters (effective temperature $\Teff$, surface gravity, metallicity \mh\ and abundance of $\alpha$-elements with respect to iron). }
{Publicly available, 3D high-resolution maps at various scales around the Sun are produced through the computation of differential extinction. This is achieved by discretising the dataset into spherical coordinates. Thanks to an all-sky coverage, our map allows the identification of all the major extinction structures. Our large scale map covers a region of 4$\times$4$\times$0.8 kpc centred on the Sun with a discretisation of ($dr, d\theta, d\phi) = (40~\text{pc}, 1\deg, 1\deg)$. To exploit the higher number of stars in the proximity of the Sun, we created a smaller scale map focused on the Local Bubble area, with a volume of 1$\times$1$\times$0.8 kpc and a finer discretisation of ($dr, d\theta, d\phi) = (30~\text{pc}, 1\deg, 1\deg)$.}
{The produced interstellar extinction maps exhibit a strong spatial correlation with known structures such as molecular clouds and the spiral arms. Interestingly, several regions of the map are consistently present in different tracers as the density of gas, of young stars and the chemical pattern of the spiral arms. In agreement with the literature, the extinction tends to be higher at positive azimuthal angles. Our study unveils the link between the Galactic distribution of dust, gas, and stars governing the chemical and dynamical evolution of the spiral arms in the Galactic ecology framework.}


\titlerunning{\Gaia\ \gspspec\ 3D extinction maps of the Milky Way disc: tracing dust–gas–star interactions}
\authorrunning{Barbillon et al.}

\maketitle\
\section{Introduction}
\label{Introduction}


The mapping of the interstellar extinction in the Galaxy traces the distribution of dust, particularly in regions of the Milky Way (MW) disc at low Galactic latitudes. High dust and gas-density areas are, in fact, expected to be associated with regions of strong star formation \citep{dacunha2010,Parente2024}. The study of the interstellar medium (ISM) in the Galactic disc, therefore, provides a better understanding of its spatial structure \citep{calura2017}. Interstellar extinction, caused by dust along the line of sight (l-o-s) scatters starlight, making it fainter and redder. This misrepresents the absolute magnitudes and intrinsic colours of the stars. 

Traditional two-dimensional (2D) extinction maps show the distribution of cumulative extinction in a given direction integrated along the l-o-s. Consequently, these extinction values represent an upper limit of the real extinction of a local disc star in that direction \citep{Schlegel1998, Planck_collab_2014_dust}. Large photometric and spectrophotometric stellar surveys have started to provide extinction and absorption data. When coupled with distances, they bring a 3D perspective of the ISM that has allowed the development of the first inversion models. These models estimate the extinction by inverting the relation between the observed light and the intrinsic properties of stars \citep{Arenou1992, Drimmel01, Vergely01, Chen13}. However, spectroscopic methods have provided a more direct approach by analysing absorption and emission features in the spectra of stars and galaxies such as diffuse interstellar bands (DIBs) or spectral lines of specific gaseous species \citep{Calzetti2000, Marshall06, vergely10, Schlafly2015, lallement_14, Schultheis2014, Rezaei2017, DR3_carto_DIB} to constrain the extinction law (\citealt{Fitzpatrick1999, Gordon2003}). Recently, the combination of \Gaia~parallaxes and information about stellar positions and velocities brought all the ingredients into the determination of individual extinctions for a huge number of stars in the Galaxy. This has led to new 3D dust maps that use a variety of procedures. Various methods, including the use of photometry with colour-colour diagrams \citep{MonrealIbero2015, anders_2022}, machine learning \citep{Chen19}, main-sequence turn-off fitting \citep{Gontcharov2017}, SED fitting with grid or MCMC sampling \citep{Berry12, Chen2014, Green14, Sale2014, Guo21}, and comparisons with synthetic stellar populations \citep{green15, green18, green19}, have been used to infer the local extinction density and derive individual stellar extinctions. Developments in modern statistical techniques have led to great strides in mapping the 3D structure of the dust in our Galaxy, especially exploiting Bayesian methods and Gaussian processes \citep{Sale_Magorrian_14, Capitanio17, lallement2019, green19, Leike20, Babusiaux2020, Vergely22, Zhang23, Edenhofer_2024}. Recently, in contrast to the inversion method, \cite{Dharmawardena24} uses a forward model to infer the 3D dust density and l-o-s extinction of any user-selected region from arbitrary catalogues of stellar extinctions, using distances and sky positions of stars as input.

As a result, each mapping technique has its own advantages and limitations. Statistical approaches are best suited for large distances as they rely on a vast number of targets, whereas combining individual l-o-s data is more efficient for regions closer to the Sun. Ground-based near-infrared (NIR) and infrared (IR) surveys can probe deeper into dense clouds than optical observations but face challenges in correcting for telluric emissions and absorptions. In addition, they are less efficient at detecting high temperature stars. 
Catalogues of targets observed in spectroscopy are considerably smaller than photometric catalogues, but purely photometric determinations of the extinction are less accurate because of the strong degeneracy between stellar temperature, metallicity, and reddening when relying only on colours band. Although observations in the IR reduce the impact of extinction, they do not completely remove this degeneracy. In this context, the \Gaia\ mission \citep{Gaia_photometry, Gaia_mission} and its DR3 \citep{DR3Gaia_2023} has reached the state-of-the-art data. It has provided 1.8 billion sources measured in photometry (G$_{BP}$, G, G$_{RP}$ bands in the optical range) and astrometry (positions, parallaxes, and proper motions). In addition, more than 33 million stars were spectroscopically observed by the radial velocity spectrograph\footnote{\url{https://www.cosmos.esa.int/web/gaia}} (RVS) in the NIR calcium triplet wavelength range \citep{Katz2023}. The RVS is an integral field spectrograph with resolving power $\sim$11500 that covers the wavelength range of 845-872 nm \citep{cropper_2018}. RVS spectra are treated by the General Stellar Parametriser from Spectroscopy (\gspspec) workflow \citep{DR3_RVS} which is one module of the Data Processing and Analysis Consortium (DPAC) coordination unit 8 (CU8) of the Apsis chain described in \cite{Creevey2023}. \gspspec\ is estimated stellar atmospheric parameters for 5.6 millions stars such as the effective temperature $\Teff$, surface gravity $log(g)$, the global metallicity\footnote{In the \gspspec~catalog, \mh\ traces the iron abundances \feh\ as iron lines dominate the non-$\alpha$ element features used in the analysis.} \mh\ and individual abundances. de Laverny et al. (in prep) estimated the reddening extinction \ebprp\ in the \Gaia~G$_{Bp}$\footnote{Hereafter $B_P$ and $R_P$, respectively.} and G$_{Rp}$$^{3}$ bands using the \gspspec~stellar atmospheric parameters. We also draw attention to a second module of the CU8 Apsis chain, the General Stellar Parameterizer from Photometry \citep[\textit{GSP-Phot};][]{andrae_2023}. This module is also designed to characterise single stars and interstellar extinction in Gaia DR3 but using the low-resolution $B_P$/$R_P$ spectra instead of the high resolution RVS spectra.

The aim of this article is to provide a detailed analysis of the extinction distribution in the MW, as a way to trace the interactions between dust, gas, and stars in the disc. More generally, our objective is to explore the extinction derived using a purely spectroscopic-based dereddened colour from the \Gaia\ \gspspec\ data; these reddening are thus fully compatible with these spectroscopic parameters. The results are used to construct 2D and 3D maps of the Galactic disc that are compared to the literature. The paper is structured as follows: In Section \ref{data_estimation}, we explain in detail the estimation of extinction with the first 2D maps for the all-sky \Gaia\ observations. Section \ref{data_selection_section} introduces the data selection of the MW disc sample and its statistical analysis. Section \ref{3Dmaps_section} describes the method to build the 3D extinction map, how to access the 3D extinction data and our results for two regions of the Solar vicinity. Then, section \ref{Discussion} details our discussion and conclusions. 

\section{Reddening estimation}
\label{data_estimation}

\begin{figure*}[htbp]
\includegraphics[width=0.95\linewidth]{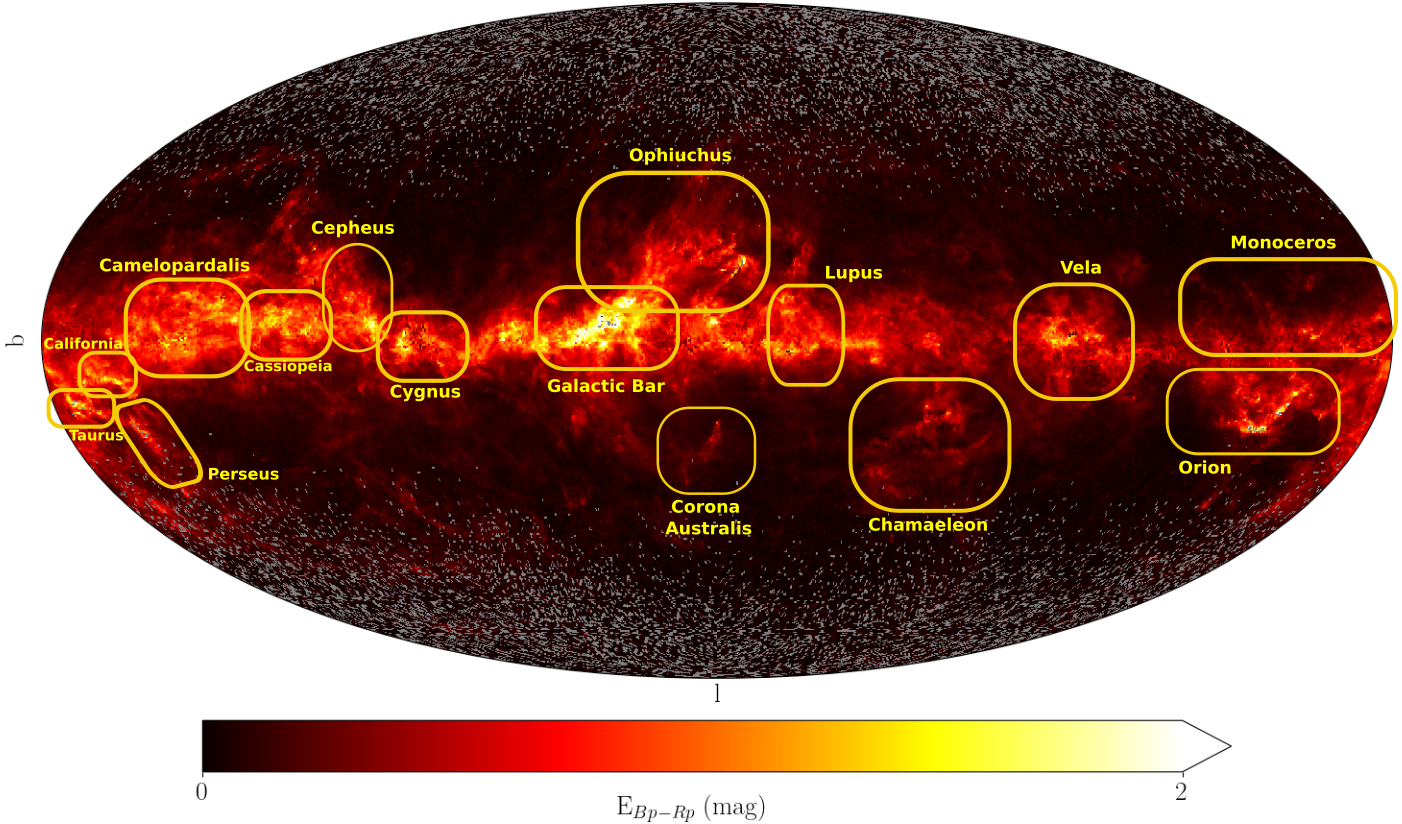}
    \caption{2D HEALPix \citep{Gorski2005} full-sky map of \ebprp~using selected star of the \Gaia~DR3 catalogue. The spatial resolution is 0.46\deg~per HEALpixel (level 7). The colour scale indicates the \ebprp~distribution. Here only 2398038 stars with $D > 600$ pc are shown. In yellow, we overplotted known interstellar dust structures from the literature.}
\label{lvsb_plot}
\end{figure*}

\begin{figure*}[htbp]
    \centering
    \includegraphics[width=0.8\textwidth]{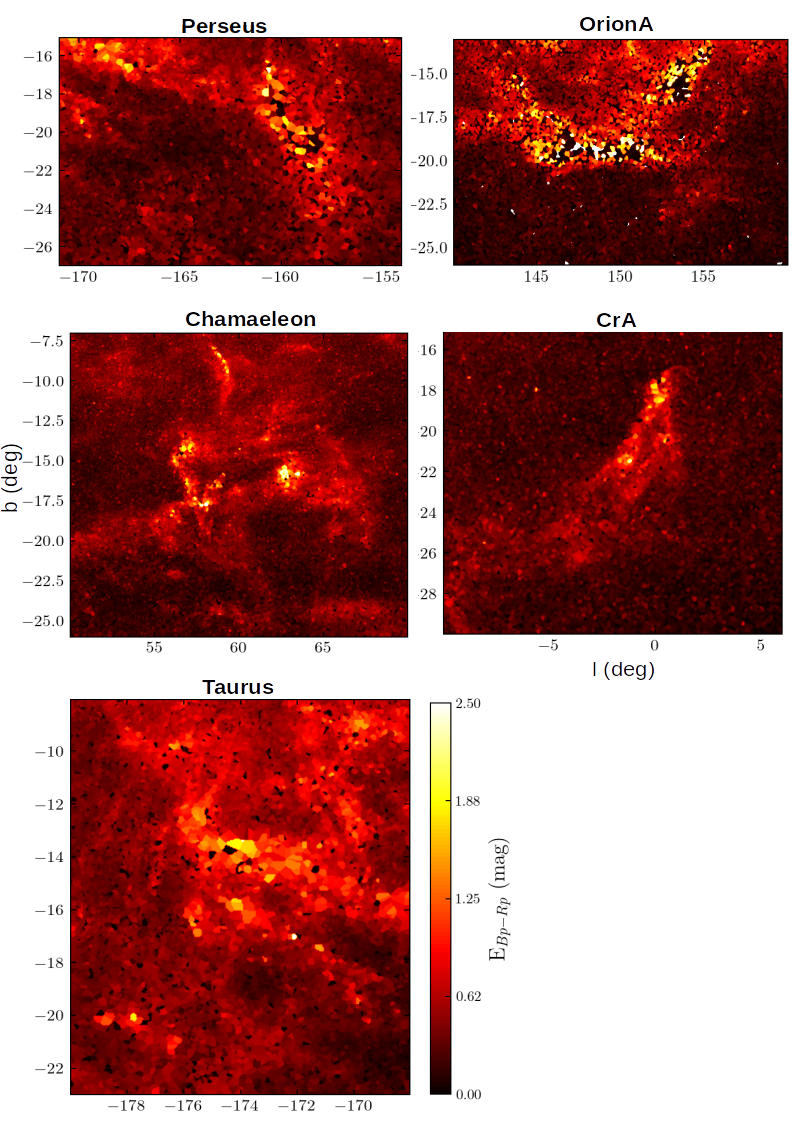}
    \caption{Zoomed-in views of Fig. \ref{lvsb_plot} toward different individual molecular clouds (Perseus, OrionA, Taurus, Corona Australis (CrA), and Chamaeleon) as presented in \cite{Edenhofer_2024}. The colour scale indicates the \ebprp~distribution.}
    \label{zoom_regions}
\end{figure*}


\begin{figure*}[htbp]
\includegraphics[width=0.95\linewidth]{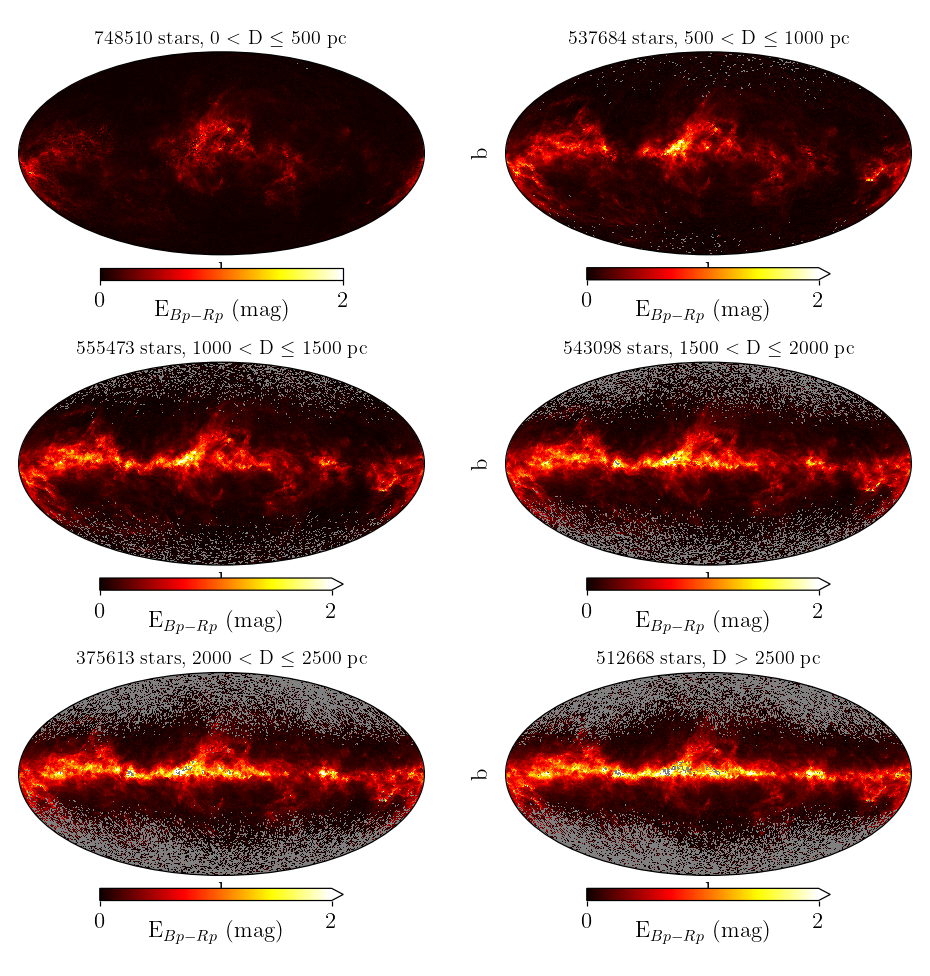}
    \caption{Same as Fig. \ref{lvsb_plot} (without excluding stars with $D>600$ pc), but separating the sky in distance intervals of 500 pc and with a spatial resolution of 0.92$\deg$ per HEALpixel (level 6).}
\label{lvsb_plot_d_bins}
\end{figure*}

Since all the atmospheric parameters, and in particular the effective temperature derived from \Gaia\ RVS data by \gspspec\ are not affected by interstellar extinction, 
de Laverny et al. (in prep.) compared the observed stellar $(B_P-R_P)_\star$ colour with the theoretical colour $(B_P-R_P)_0$ to infer the interstellar reddening. To this purpose, the theoretical $\Teff$-log($g$)-\mh–colour relation of \cite{Casagrande2021} (see their Equation 1 and coefficients from Table 1) has been used considering the stellar atmospheric parameters derived by \gspspec. 
The \ebprp\ can simply be derived by subtracting the theoretical colour to the observed one. To account of the errors, de Laverny et al., in prep, performed 1000 Monte Carlo realisations of the \ebprp\ considering the uncertainties of the stellar atmospheric parameters and observed colour. From these Monte Carlo realisations, the 84$^{th}$ quantiles (upper limit) and 16$^{th}$ quantiles (lower limit) are provided as indicators of the uncertainty. Furthermore, a quality flag (flag\_Redd) to select the best reddening estimations is available. This flag can take the values of 0, 1 or 2, ranging from high to moderate quality. We refer to de Laverny et al. (in prep.) for more details on the adopted procedures and on the definition of this quality flag. Notably, in DR3 data, a $\Teff$ limitation of the \gspspec\ reference spectra grid\footnote{This grid will be updated (or extended) in DR4.} must be considered, restricting the studied stars to $\Teff$ cooler than 8000 $K$. 

Combining the galactic longitude $\ell$ and the Galactic latitude $b$ from \Gaia\ astrometry, we can create 2D all-sky map. Although, the entire catalogue can be used to create these 2D maps, we applied some cuts to avoid poor measurements that are particularly related to distant stars at high or low latitudes. We considered the specific flags as follows:

\begin{itemize}
    \item[$\circ$] flags\_gspspec < 2
    \item[$\circ$] flag\_Redd = 0
\end{itemize}   

where flags\_gspspec are the first thirteen atmospheric parameter flags from Table 2 of \cite{DR3_RVS} and the flag\_Redd is the quality flag associated to the reddening calculation (cf. Table 1 from de Laverny et al., in prep). With these criteria, the \gspspec\ catalogue decreases to 3273046 number of stars.

Figure \ref{lvsb_plot} shows a 2D all-sky map of the extinction \ebprp\ distribution of each stars of the \gspspec\ catalogue considering the previous mentioned flags. As the number of the stars is increasing in the vicinity of the Sun, we avoid the regions where $D<600$ pc for visualisation purposes, to avoid the closest stars with very low extinction that dominate the line of sight. As expected, larger extinction values in the Galactic plane are observed. Our map exhibits strong overall agreement with existing data. Known structures are clearly recovered and comparable to 2D maps created from the thermal emission of interstellar dust detected in the IR \citep{Schlegel1998}, photometric surveys \citep{Marshall06, green19, Dharmawardena24} or a combination of spectroscopic and photometric data \citep[][and references therein]{lallement_2018, Chen19, Zhang23, Edenhofer_2024}. The consistency between the various 2D maps presented in the literature remains imperfect, highlighting the need for further studies to refine them.
As an example, \cite{Edenhofer_2024} (cf. their Section 7.1, Fig. 11) presents a comparison of Mollweide projections of total integrated extinction from different studies \cite[][]{Planck_collab_14, green19, Leike20, Leike22, Vergely22, delchambre_2023}. Every 2D all-sky map and Fig. \ref{lvsb_plot} consistently reveal fine structures at high Galactic latitudes. However, divergences arise in the Galactic plane due to variations in distance estimation, probed spatial volume and the different techniques involved to estimate extinction. For instance, the standard deviation of the extinction between -10<$b$<10° is equal to 0.24 mag, 3.29 mag, 0.20 mag and 0.29 mag for our converted \ebv\ and those of SFD, \citep{Edenhofer_2024} and Bayestar19, respectively. Interestingly, Figure 12 of \cite{Edenhofer_2024} presents a zoomed-in view towards five individual molecular clouds seen in the 2D all-sky map, comparing different literature works. Our Fig. \ref{zoom_regions} reproduces similar zoomed-in views from our data for the Perseus (left upper panel), OrionA (right upper panel), Taurus (left middle panel), Corona Australis (CrA) (right middle panel) and Chamaeleon (bottom panel) molecular clouds. Comparisons of our \Gaia\ extinction zoomed-in maps show sharp consistency with the literature. For illustration purposes, the maps have been smoothed using a Gaussian bivariate kernel density estimation (KDE) \citep{Feigelson2012}. To visualise the different extinction structures, we optimise the bandwidth $h$ depending on the area of the sky. In each panel, the applied bandwidth is $h=0.01$, $h=0.005$, $h=0.01$, $h=0.03$ and $h=0.03$ kpc, respectively. 



In addition, the \Gaia\ astrometry data provide accurate distances. Using the geometric distances from \cite{Bailer-Jones_distances}, Fig. \ref{lvsb_plot_d_bins} shows the 2D all-sky map of the extinction \ebprp\ of individual stars of the sample for several distance bins of 500 pc width. As in the case of Fig. \ref{lvsb_plot}, higher extinction values are observed in the Galactic plane. These maps reveal known structures offering novel insights into the spatial distribution of extinction across the Galactic plane.

\section{Disc sample used for the 3D extinction maps}
\label{data_selection_section}

\begin{figure*}[htbp]
    \centering
        \includegraphics[width=0.95\linewidth]{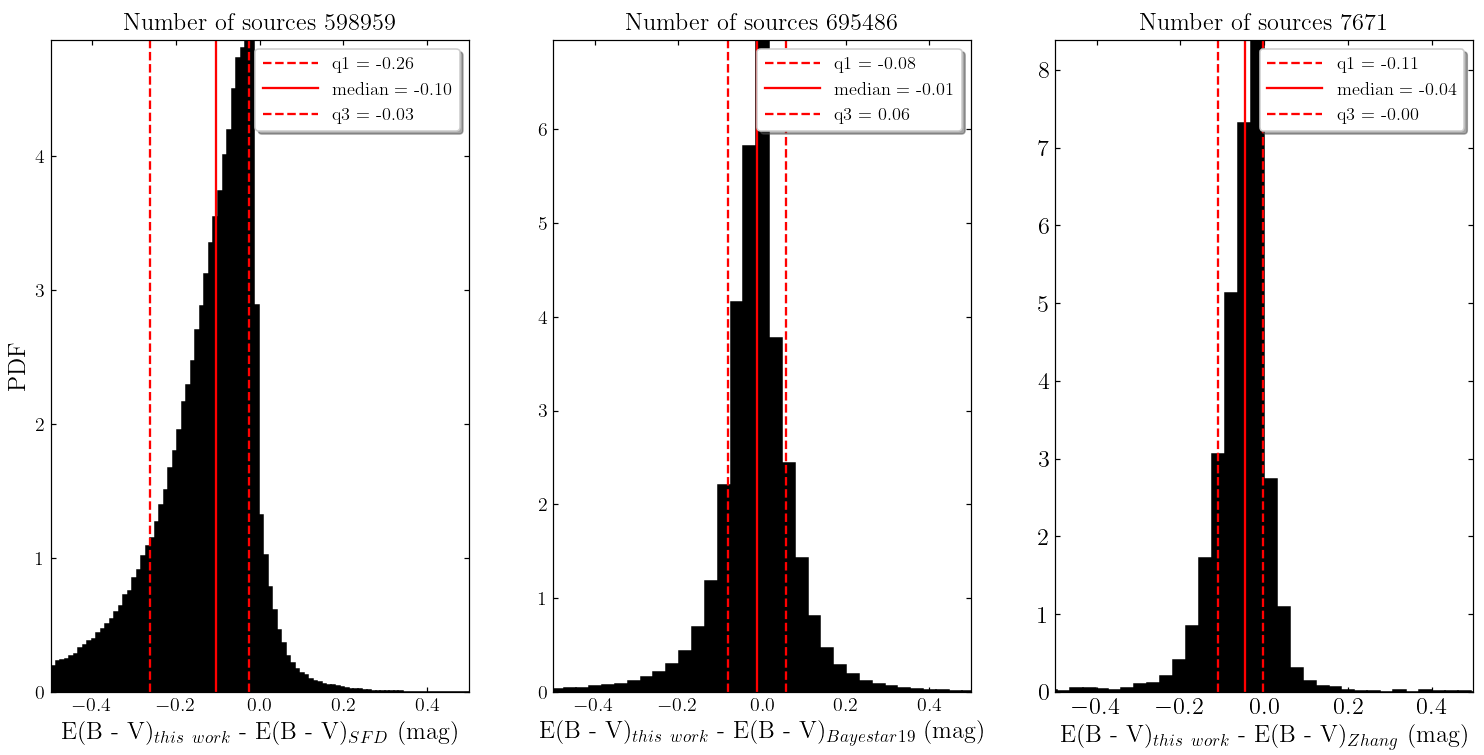}
        \caption{Comparative histograms of the difference between the converted \ebprp\ values derived in the current work with the \ebv\ values from \cite{Schlegel1998} (left panel), \cite{green19} (middle panel) and \cite{Zhang23} (right panel). Red straight lines denotes the median difference while the red dashed lines are the first and third quartiles, respectively.}
    \label{comparaison_hist_extinction}
\end{figure*}

\begin{figure*}
    \centering
        \includegraphics[width=0.95\linewidth]{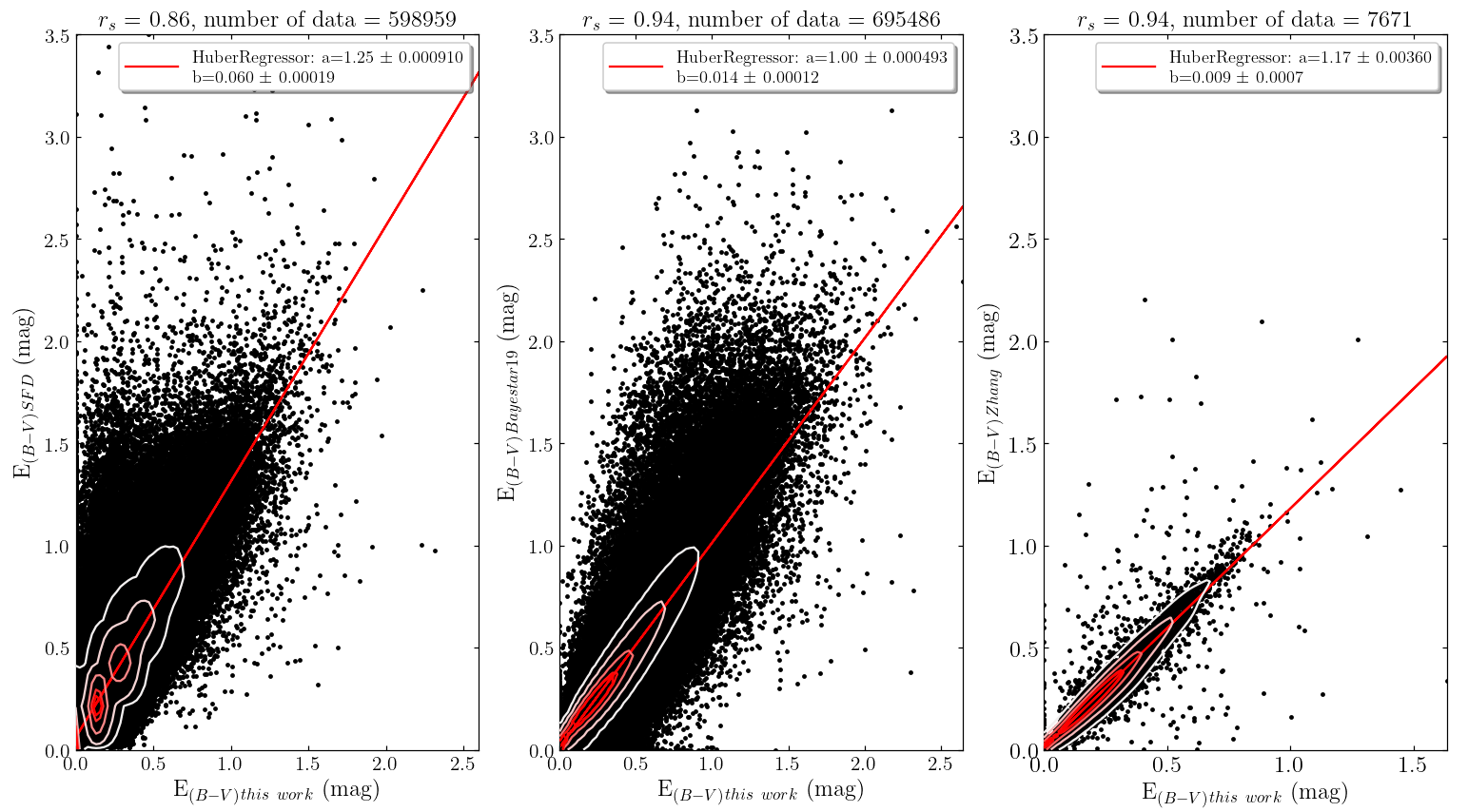}
        \caption{Correlation between the converted extinction of this article and \ebv\ values from \cite{Schlegel1998} (left panel), \cite{green19} (middle panel) and \cite{Zhang23} (right panel). The white contour lines enclosing fractions of 90, 75, 60, 45, 30, and 20\% of the total number of stars. The red solid line represents the Huber loss function, a linear regression model (with $a$ being the slope slope and $b$ the intercept) that is robust to outliers with $\epsilon$=1.35 (default value). This parameter $\epsilon$ controls the number of samples that should be classified as outliers. The spearman correlation coefficient $r_s$, the number of considered targets, and the fitting gradients are indicated on the top of each plot.}
    \label{comparaison_extinction}
\end{figure*}

To obtain the 3D extinction maps, we performed different quality selections on the \Gaia\ data, the distances and the derived \ebprp. The Cartesian coordinates $X,Y,Z$ are retrieved by combining the \Gaia\ $\ell$ and $b$ coordinates, the geometric distances of \cite{Bailer-Jones_distances}, the Sun's Galactocentric distance $\Ro=8.249$ kpc from \cite{gravity_collab_2021} and the height above the Galactic plane Z$_{\odot}$ = 0.021 kpc \citep{Bennett_2019}. The defined sample is restricted as follows by excluding poor measurements and focusing the MW disc:

\begin{itemize}
    \item[$\circ$] flags\_gspspec < 2
    \item[$\circ$] flag\_Redd = 0
    \item[$\circ$] relative error : \ebprp$_{unc}$/\ebprp\ < 10\%
    \item[$\circ$] ruwe < 1.4
    \item[$\circ$] relative error : d$_{geo_{unc}}$/d$_{geo}$ < 10\%
    \item[$\circ$] -0.4 < Z < 0.4 kpc

\end{itemize}   

where flags\_gspspec are the first thirteen atmospheric parameter flags from Table 2 of \cite{DR3_RVS}, ruwe is the renormalised unit weight error from \cite{Lindegren2018} and the flag\_Redd is the quality flag associated to the reddening calculation (cf. Table 1 from de Laverny et al., in prep). Uncertainties of distances and extinctions are defined as half of the difference between its upper and lower confidence levels 
(de Laverny et al. in prep).
With these criteria, the disc sample of this article contains 1118507 stars, characterised by an observational volume of 8$\times$8$\times$0.8 kpc centred at the Sun. Figures \ref{comparaison_hist_extinction} and \ref{comparaison_extinction} compare our extinction values of the individual stars with previous studies \citep{Schlegel1998, green19, Zhang23}. For this purpose, we first converted the \ebprp\ into E$(B-V)$. We assumed that the extinction curve $R(\lambda)$ is universal for all stars. That is, the extinction of any given star is proportional to a single, universal function of wavelength: $A(\lambda) \propto R(\lambda)$\footnote{$A(V) = R_V \times \ebv$.}, assuming $R_V=3.1$ \citep{Cardelli1989}. We estimated $A(B_p)/A(V)$ and $A(R_p)/A(V)$ from the photometric system of the ESA/Gaia website\footnote{\url{http://stev.oapd.inaf.it/cgi-bin/cmd_3.8}.} selecting the \Gaia\ EDR3 (all Vegamags, \Gaia\ passbands) and the \textit{YBC+new Vega version}: $A(B_p)/A(V)$=1.08337 and $A(R_p)/A(V)$ = 0.63439. We know that this relation does not account for the detailed variations in stellar atmospheric parameters (particularly $\Teff$ in our case). However, we verified whether any systematic trend with effective temperature and extinction was present, and none was found. In summary, for each star the conversion has been computed using the following relation: 

\begin{align}
\begin{split}
    E(B_p - R_p) =&~(B_p - R_p)_{\star} - (B_p - R_p)_0 \\
    &=(1.08337 - 0.63439) \times A(V) \\
    &=(1.08337 - 0.63439) \times R_V \times \ebv
\end{split}
\end{align}

where $(B_p - R_p)_{\star}$ and $(B_p - R_p)_0$ are the measured colour index and intrinsic colour index, respectively.

Globally, our results show strong consistency with the literature, validating our approach. Fig. \ref{comparaison_hist_extinction} presents histograms of the \Gaia~\gspspec~extinction variation compared to the literature. Firstly, \cite{Schlegel1998}\footnote{The extinction catalogue was generated from Python dustmaps package \url{https://dustmaps.readthedocs.io/en/latest/index.html}.} (hereafter SFD) derived all-sky 2D reddening maps based on far-infrared emission from interstellar dust observed by the IRAS mission \citep{Neugebauer1984} and the DIRBE instrument on COBE missions \citep{Boggess1992}. They estimated the dust temperature from the ratio of 100 $\mu m$ and 240 $\mu m$ DIRBE fluxes, and combined this with high resolution IRAS 100 $\mu m$ maps to infer the density of the dust column. This emission was then calibrated against the colours of elliptical galaxies to obtain $E(B - V)$ values. However, due to IRAS flux saturation and contamination by infrared cirrus, regions within $|b| < 5^\circ$ were excluded from their analysis and, therefore, from our comparison. We obtain a median difference of -0.10 mag, with a first quartile equal to -0.26 mag and a third quartile equal to -0.03 mag. Although regions near the Galactic plane were excluded, some contamination from stars affected by high extinction remains, which may explain the large values. In addition, SFD gives only the total extinction for a given line of sight and therefore do not deliver the information of dust distribution as a function of distance. It explains this slightly high median in the negative values and the extinction values higher than ours. Then, \cite{green19}$^{8}$ (hereafter Bayestar19) derived the extinction maps using a Bayesian inference method with a prior on the distribution of interstellar dust. To this purpose, they combined the photometry and the stellar properties from \Gaia, Pan-STARRS1, and 2MASS surveys for about 799 million stars. Bayestar19 covers the southern sky only for declination $\delta$ > -30\deg, consequently, we limited our comparison to the same spatial coverage. We obtain an excellent agreement, quantified with a median difference of -0.01 mag, with a first quartile equal to -0.08 mag and a third quartile equal to 0.06 mag. Finally, \cite{Zhang23}\footnote{Data published online \url{https://zenodo.org/records/7811871}.}  derived the extinction for 220 millions of stars. Their data-driven procedure combines the \Gaia\ $BP/RP$ spectra ($R \sim$ 30–100) with infrared photometry from 2MASS and unWISE \citep{Schlafly_2019}. To create our comparison catalogue, we considered only stars matching their "basic reliability cut", i.e. stars with quality flags < 8. We again obtain an excellent agreement with a median difference of -0.04 mag, with a first quartile equal to -0.11 mag and a third quartile equal to 0.00 mag. 
Fig. \ref{comparaison_extinction} shows the correlation between our E($B-V$) and the three works of literature mentioned above. The white contours show the fraction distribution of each sample. The solid red line represents the Huber loss function, a linear regression model that is robust to outliers with a fixed $\epsilon$=1.35 (default value). This parameter $\epsilon$ controls the number of samples that should be classified as outliers. In each case, the slope is positive and close to 1, with 1.25 $\pm$ 0.000910, 1.00 $\pm$ 0.000493 and 1.17 $\pm$ 0.00360 for SFD, Bayestar19 and Zhang, respectively. For the case of SFD, the slope is slightly higher due to the persistence of some contaminated stars affected by strong extinction in the middle plane and because their extinction is a cumulative one. In each case, the spearman correlation coefficients $r_s$ \citep{Spearman1904} are high equal to 0.86, 0.94 and 0.94 for SFD, Bayestar19 and Zhang, respectively.

\section{3D extinction maps}
\label{3Dmaps_section}

3D maps of Galactic interstellar dust are a useful tool for a wide range of purposes, such as investigating the morphology of ISM in the MW, to constraining star formation regions or improving distance or luminosity estimate through extinction corrections (cf. Sect. \ref{Introduction}). 
To build the 3D maps, we used the sample detailed in Sect. \ref{data_selection_section}, then computed the differential extinction along the l-o-s, $\Delta$E. To this end, we discretised the data by binning in spherical coordinates ($r$, $\phi$, $\theta$) to divide the space into quasi-cubic volumes. Along each radial l-o-s, we then calculated $\Delta$E:
\begin{equation}
    \Delta E = E_{n+1}-E_{n}
\end{equation}

where $E_{n}$ is the median extinction in the $n$-th volume with $n$ increasing with distance. For robustness, only volumes that included more than 2 stars were considered. The number of targets is limited beyond 2 kpc in the Galactic plane (cf. Fig. \ref{stat_3Dmaps_KDE_sunvicinity} that illustrates the 3D distribution of the number of stars). Consequently, we refined the initial observational volume of 8$\times$8$\times$0.8 kpc to a smaller one of 4$\times$4$\times$0.8 kpc centred on the Sun to create a \textit{large scale} extinction map (see Sect. \ref{3Dmaps_section_sunvicinity}). The discretisation applied in the spherical coordinates is ($dr, d\theta, d\phi) = (40~\text{pc}, 1\deg, 1\deg)$. 

To exploit the higher number of stars in the proximity of the Sun, we created a smaller scale map focused on the Local Bubble (LB) area. To this end, we used an observational volume of 1$\times$1$\times$0.8 kpc centred on the Sun. This second scale is referred as the \textit{small scale} maps in Section \ref{LB_section}. The discretisation applied in the spherical coordinates is ($dr, d\theta, d\phi) = (30~\text{pc}, 1\deg, 1\deg)$, yielding a higher resolution than the \textit{large scale} maps. 


The 3D extinction data are publicly available in Zenodo at \url{https://doi.org/10.5281/zenodo.17598127}. We shared a code to reproduce the 2D projected maps and apply the Gaussian KDE discussed in this article. Additionally, we also provided similar data using the absorption in the $G$-Band of \Gaia, $A_g$ (see de Laverny et al., in prep., for more details on this parameter).


\begin{figure*}[htbp]
\includegraphics[width=0.95\linewidth]{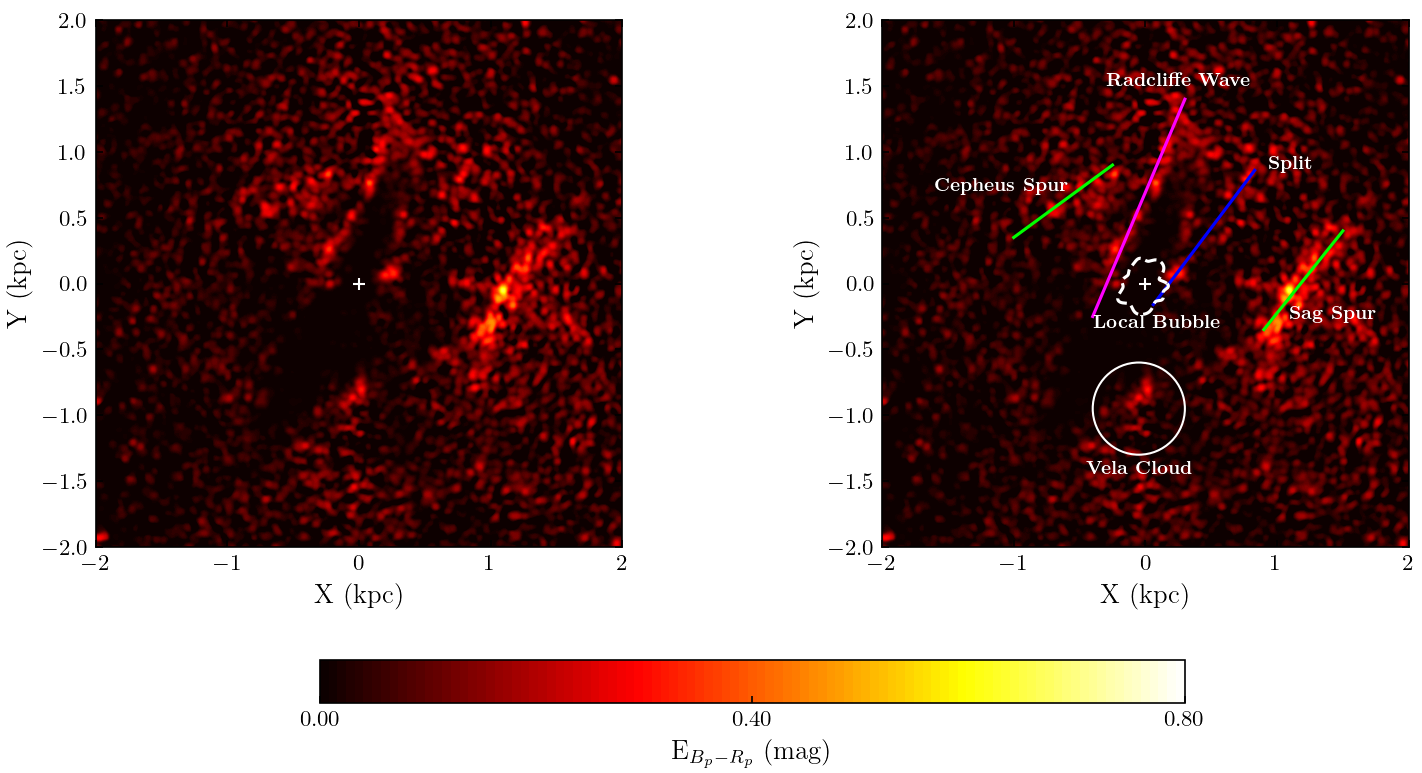}
\caption{The left and right panels show the 2D Cartesian projections of the 3D dust distribution in the Galactic disc for the $XY$ plane. The right panel includes some known structures in the proximity of the Sun. The Sun is at coordinates ($X$, $Y$) = (0, 0) kpc and is shown with a white cross. The Galactic centre direction is to the right. The brightest regions correspond to the densest structures and cloud cores, as opposed to the darkest regions. The Radcliffe Wave, the Split and the Cepheus and Sag spurs are indicated in magenta, in blue and in green, respectively. The dashed white line is the Local Bubble contours \citep{LB_Pelgrims20}. The discretisation in the radial direction $dr$=40 pc and d$\theta$=d$\phi$=1\deg. The bandwidth of the KDE is $h$ = 20 pc.}
\label{3Dmaps_KDE_sunvicinity_XY}
\end{figure*}

\begin{figure*}[htbp]
\includegraphics[width=0.90\linewidth]{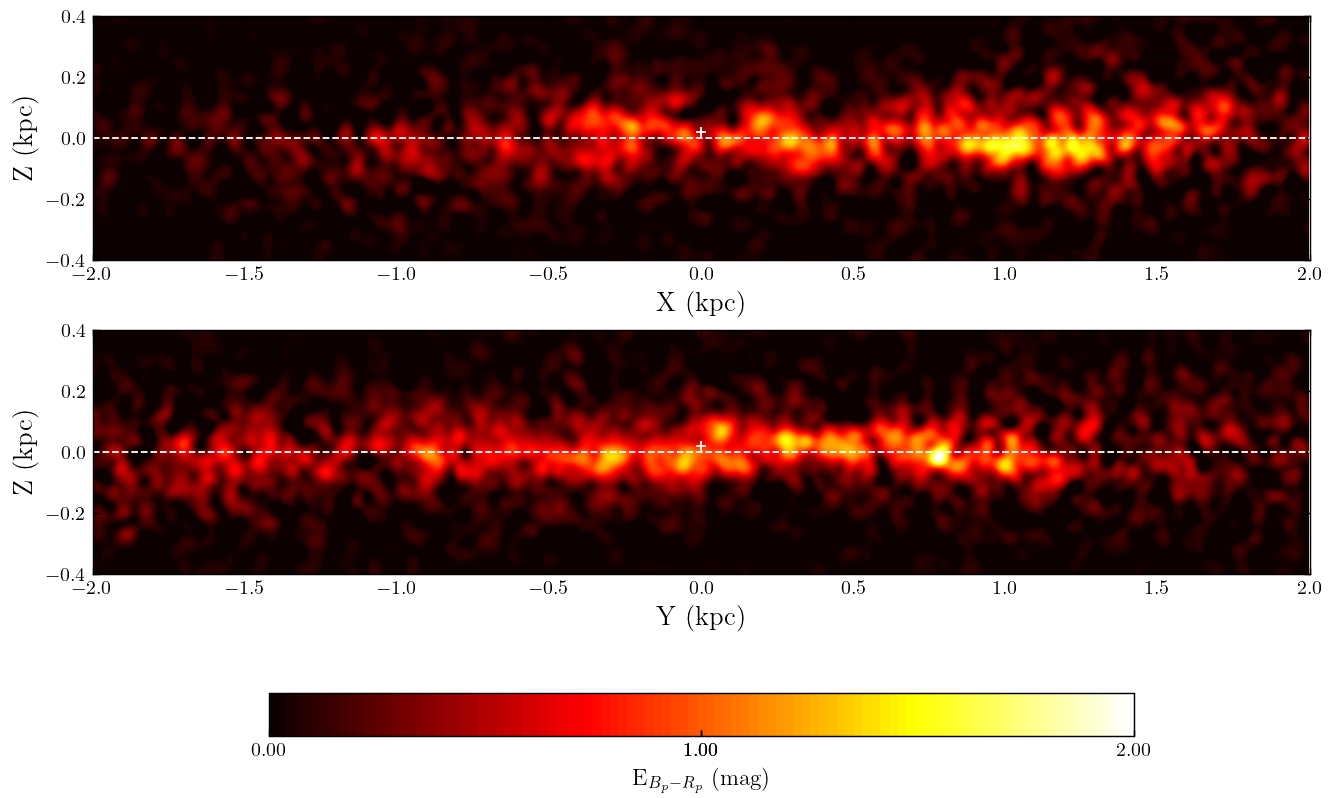}
\caption{Same as the Fig. \ref{3Dmaps_KDE_sunvicinity_XY}, but for the $XZ$ (upper panel) and $YZ$ (bottom panel) planes. The Sun is at coordinates ($X$, $Z$) = ($Y$, $Z$) = (0, 0.021) kpc. The dashed white line represents the galactic plane at $Z=0$.}
\label{3Dmaps_KDE_sunvicinity_Zplane}
\end{figure*}

\subsection{Large scale maps}
\label{3Dmaps_section_sunvicinity}

Figure \ref{3Dmaps_KDE_sunvicinity_XY} presents the 2D projections of the \ebprp\ distribution in the Galactic disc by accumulating the extinction through 0.8 kpc in the $Z$ direction. For illustration purposes, the map has been smoothed using a Gaussian bivariate KDE with a bandwidth $h=20$ pc. The left panel displays sharp and large dust groups. The right panel includes some known structures such as the LB around the Sun in dashed lines. These white contours show the inner surface of the LB from \cite{LB_Pelgrims20} model. This local cavity, is a $\sim$100 pc radius cavity of hot plasma region supposed to be created by supernova explosions that occurred over the past $10-15$ Myr \citep{LB_Maiz01, LB_Breitschwerdt16}. It is surrounded by a shell of cold, dusty gas. As expected, the area delimited by the \cite{LB_Pelgrims20} model appears to be completely dust-free (cf. Section \ref{LB_section} for more details). In addition, we can identify large-scale gaseous structures such as the Radcliffe wave (RW) \citep{Alves_2020} in magenta and the Split \citep{lallement2019} in blue, which are part of the Gould Belt\footnote{The Gould’s Belt model is a disc of young stars, gas, and dust, tilted by about 20\deg\ with respect to the Galactic plane, which has long shaped our understanding of the architecture of the local ISM.} model \citep{Zucker_2022}. The RW is a filament of gas that spans 2.7 kpc, representing the densest dust region of the MW Local arm with a wave-like shape \citep{Konietzka2024}. The Split is a gaseous structure extending over 2 kpc within the Galactic disc. In green, some spurs are also observable, such as a part of the Cepheus spur that extends from ($X$, $Y$) = (-1.0, 0.35) to (-0.25, 0.90) kpc \citep{Pantaleoni2021, Kormann2025}, and the Sagittarius spur that extends from ($X$, $Y$) = (0.90, -0.35) to (1.5, 0.4) kpc \citep{Kuhn2021, Kormann2025}.

\begin{figure}[htbp]
\includegraphics[width=1\linewidth]{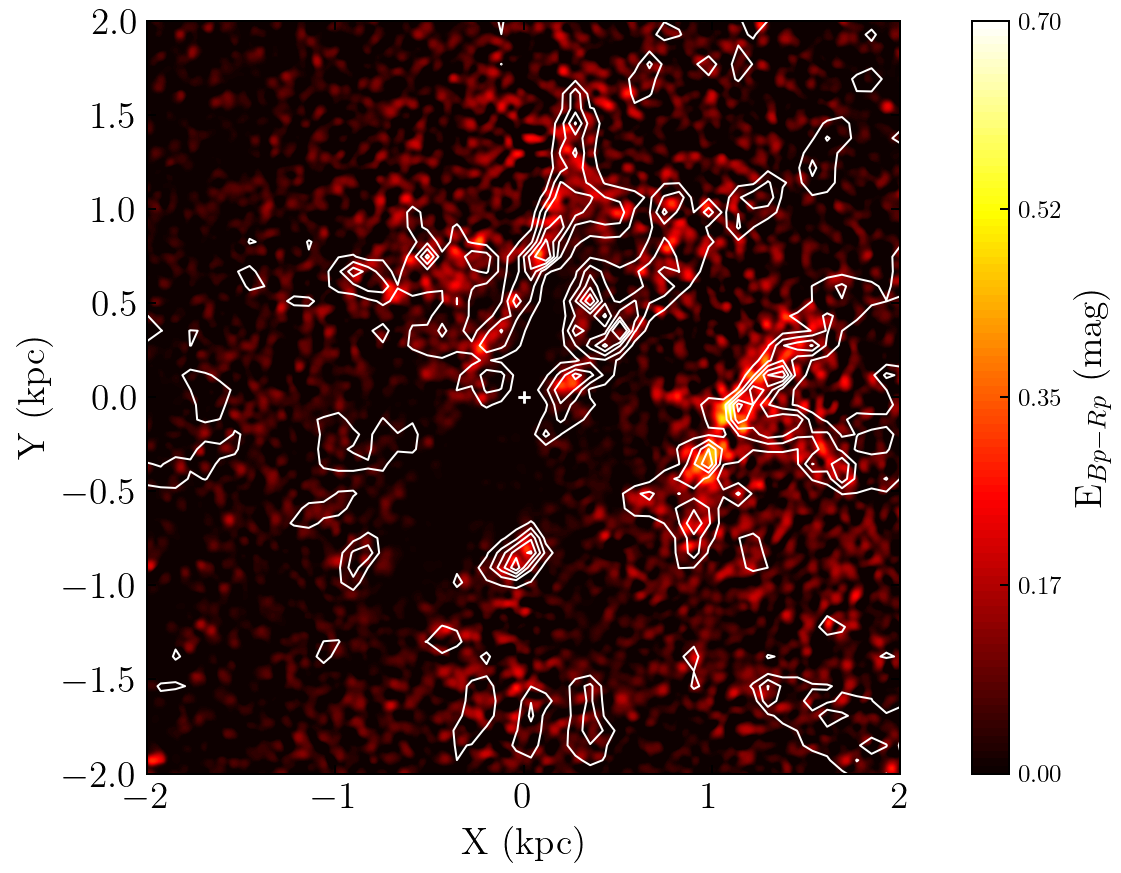}
\caption{Same as Fig. \ref{3Dmaps_KDE_sunvicinity_XY}, but with \cite{Vergely22} density contours of their dust density distribution.}
\label{3Dmaps_KDE_sunvicinity_vergelyct}
\end{figure}

Figure \ref{3Dmaps_KDE_sunvicinity_Zplane} presents the 2D projections of the \ebprp\ distribution in the Galactic disc by accumulating the extinction through 4 kpc in the $Y$ direction (upper panel) and 4 kpc in the $X$ direction (bottom panel). The upper panel of Fig. \ref{3Dmaps_KDE_sunvicinity_Zplane} shows the inhomogeneous dust distribution with respect to the Galactic plane along the $X$ axis. Interestingly, the Sagittarius-Carina region, dominating the extinction around $X$=1 kpc, is mainly located below the Galactic plane. Indeed, the extinction is increasing at larger $X$ positive values that correspond to regions closer to the Galactic centre.
The bottom panel of Fig. \ref{3Dmaps_KDE_sunvicinity_Zplane} shows a wavy pattern\footnote{This feature is even clearer in the 2D projection focused on the LB (see Sect. \ref{LB_section}).} along the $Y$-axis. The location of the higher extinction regions changes from below to above the plane when moving from negative to positive $Y$ values. This is in agreement with several independent works in the literature using different surveys and methods \citep[][]{Zari18, Alves_2020, Vergely22, Dharmawardena24, Edenhofer_2024}. For instance, Fig. 8 of \cite{Edenhofer_2024} also reveals this wavy pattern for different extinction catalogues. It seems to disappear at large distances along the $\pm Y$ axes. \cite{Edenhofer_2024} (cf. their Fig. 9) compared their reconstructed extinction to several 3D dust maps of the literature up to a distance of 1.25 kpc from the Sun. All 2D Cartesian projections maps agree on the general structure of the dust distribution despite slight differences due to the multiple methods used. Fig. \ref{3Dmaps_KDE_sunvicinity_vergelyct} shows a comparison between the \citep{Vergely22} dust density contours and our $XY$ projected extinction. The contours visually align those of this work. In addition, an agreement is also retrieved in low-extinction regions such as the giant oval cavity \citep[][cf. Figure 11]{Vergely22} in the upper left corner of our Fig. \ref{3Dmaps_KDE_sunvicinity_vergelyct}. As of the LB, the origin of this cavity would be a long series of supernovae explosions. 

\begin{figure}[htbp]
\includegraphics[width=1\linewidth]{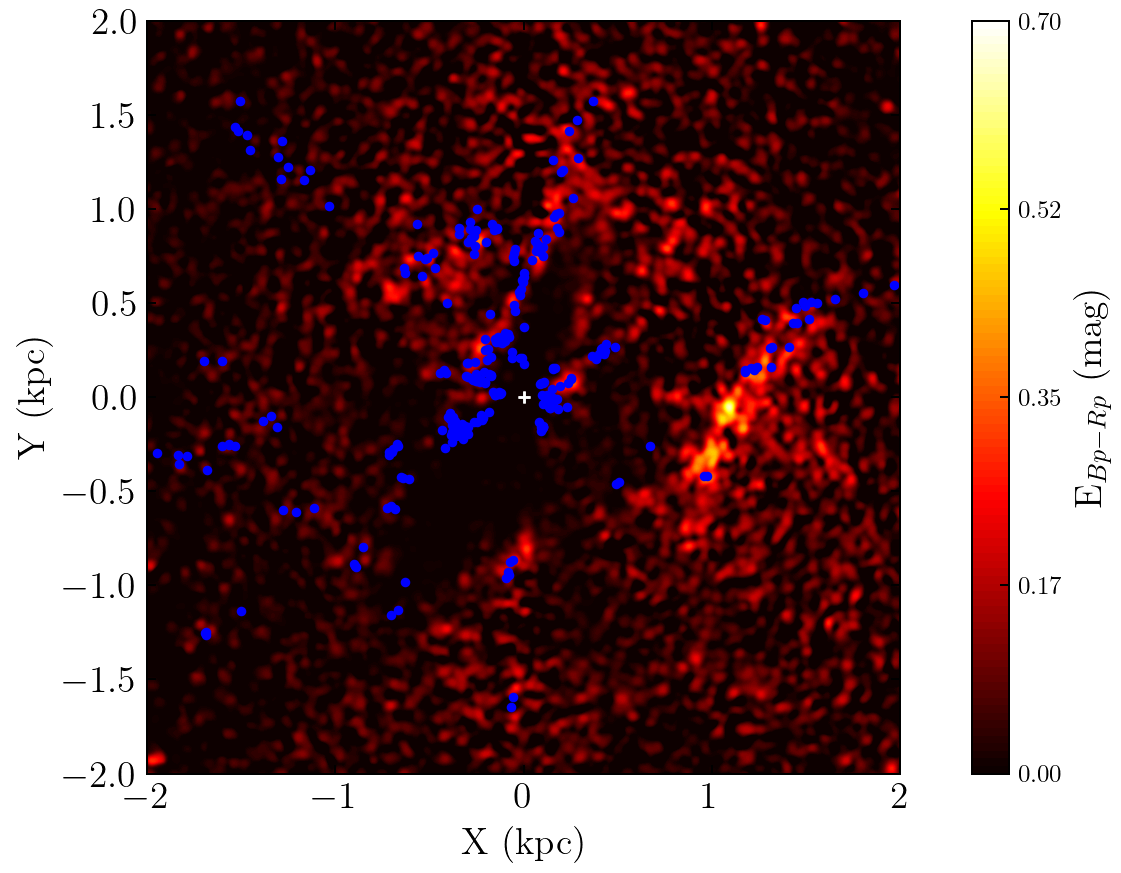}
\caption{Same as Fig. \ref{3Dmaps_KDE_sunvicinity_XY}, but with molecular clouds in blue from Star Formation Handbook catalogue \citep{Zucker2020}.}
\label{3Dmaps_KDE_sunvicinity_YSO_cat}
\end{figure}

Finally, Fig. \ref{3Dmaps_KDE_sunvicinity_YSO_cat} takes advantage 
of the most important molecular cloud catalogue 
from the Star Formation Handbook \citep{Reipurth_2008_I, Reipurth_2008_II} corrected by \cite{Zucker2020}\footnote{\url{https://dataverse.harvard.edu/dataset.xhtml?persistentId=doi:10.7910/DVN/07L7YZ}.} to perform a comparison with our dust distribution. As expected, the molecular clouds are located along the main dust features, within the reported distance uncertainties. It is interesting to note that the substructures around the LB are detected with better details in our high resolution \textit{small scale} map, presented in the following.


\subsection{Local Bubble (small scale) maps}
\label{LB_section}

\begin{figure}[htbp]
\includegraphics[width=0.98\linewidth]{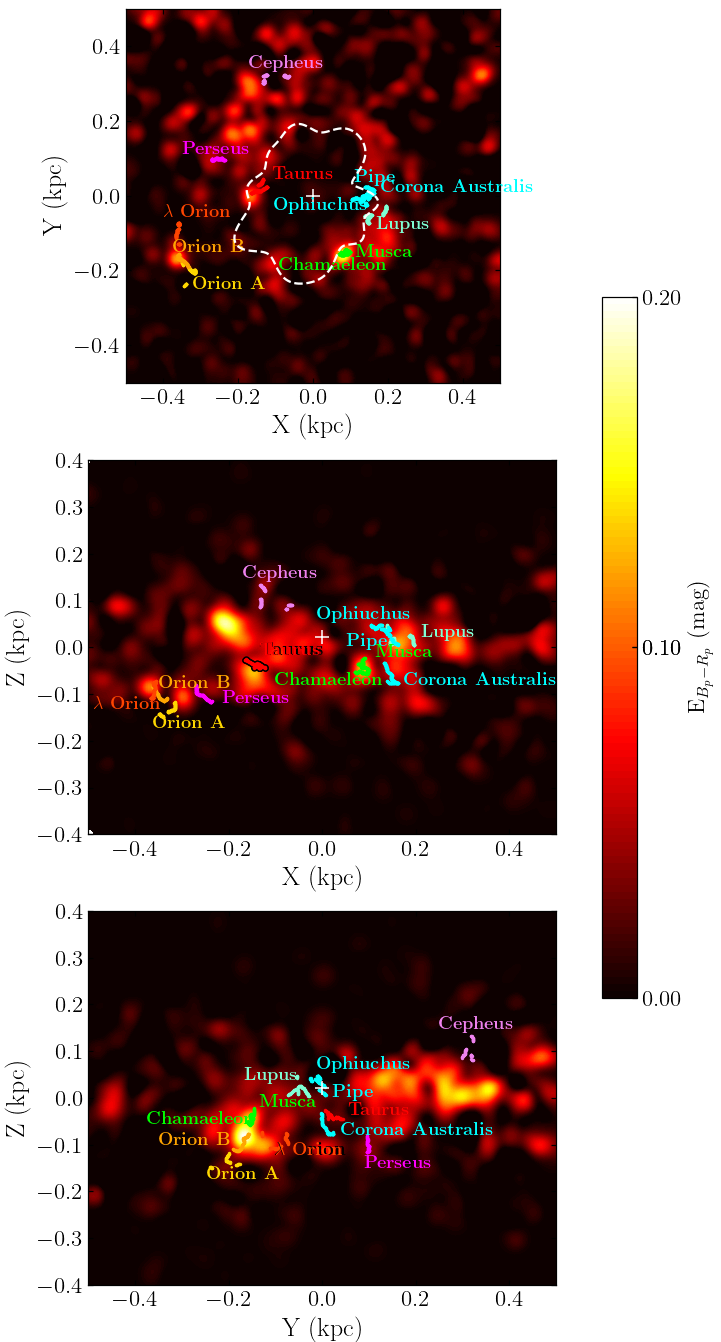}
\caption{2D Cartesian projections of the 3D dust distribution in the Galactic plane focused on the LB region. The Sun is at coordinates ($X$, $Y$, $Z$) = (0, 0, 0.021) kpc. The brightest regions correspond to the densest structures and cloud cores, as opposed to the darkest regions. Several known molecular clouds are indicated on the maps \citep{Zucker_2021, Zucker_2022}. The discretisation applied is $dr$=30 pc and d$\theta$=d$\phi$=1\deg. The bandwidth is 16 pc as the 2D grid used to cumulate the extinction in the studied direction. The upper panel is focused on the $XY$ plane with the Galactic centre direction is to the right. The dashed white lines illustrate the Local Bubble contours \citep{LB_Pelgrims20}. The middle and bottom panels show the $XZ$ and $YZ$ plane, respectively.}
\label{LB_3D_map}
\end{figure}

Figure \ref{LB_3D_map} presents the 2D projections of the \ebprp\ distribution in the LB area by accumulating the extinction through 0.8 kpc in the $Z$ direction (upper panel), 1 kpc in the $Y$ direction (middle panel) and 1 kpc in the $X$ direction (bottom panel). We have set the $h$ bandwidth to $h$=16 pc. In the left panel, the white dashed line represents the LB contours. They enclose the expansion shell driven by the supernova explosions that produced the LB. Its current shape and size have been estimated by several authors from 3D maps of the dusty ISM surrounding the Sun \citep[e.g.][and references therein]{LB_Welsh10, Capitanio17, LB_Leike19, Chen19, lallement2019}. From X-ray emission data, \cite{LHB_Liu17} confirmed its shape showing good agreement between their Local Hot Bubble (LHB) contours and the structure of the local cavity measured from dust and gas. It was later revised by \cite{LB_Pelgrims20}\footnote{Using 3D dust extinction maps of the local ISM and expended with analytic magnetic field model.}, given the current geometry of the LB contours used in this article. Despite noticeable differences in the details between \cite{LHB_Liu17} and \cite{LB_Pelgrims20} contours, the overall agreement of the LB shapes\footnote{Inferred from extinction maps and from X-ray emission.} data lends credence to the multiphase view of this local cavity. 

\begin{figure*}[htbp]
\begin{subfigure}{0.5\textwidth}
\includegraphics[width=0.95\linewidth]{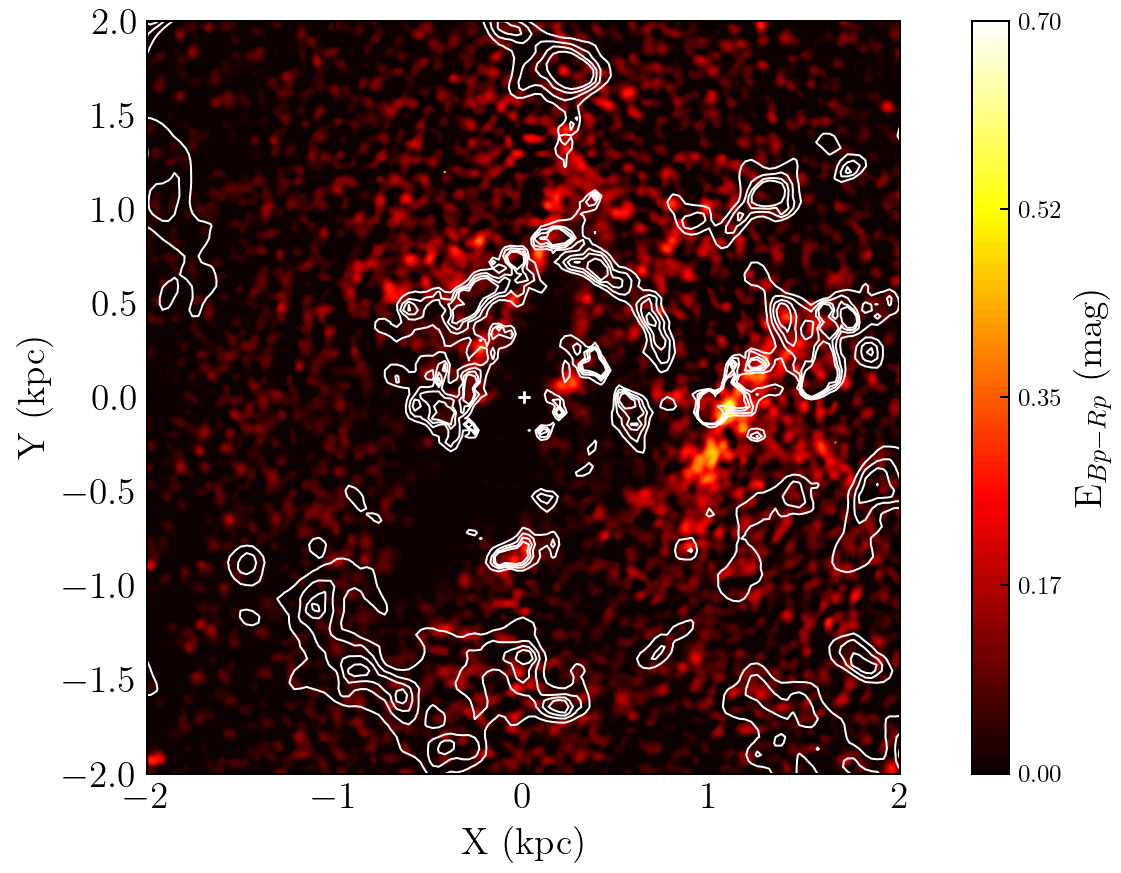}
\end{subfigure}
\begin{subfigure}{0.5\textwidth}
\includegraphics[width=0.95\linewidth]{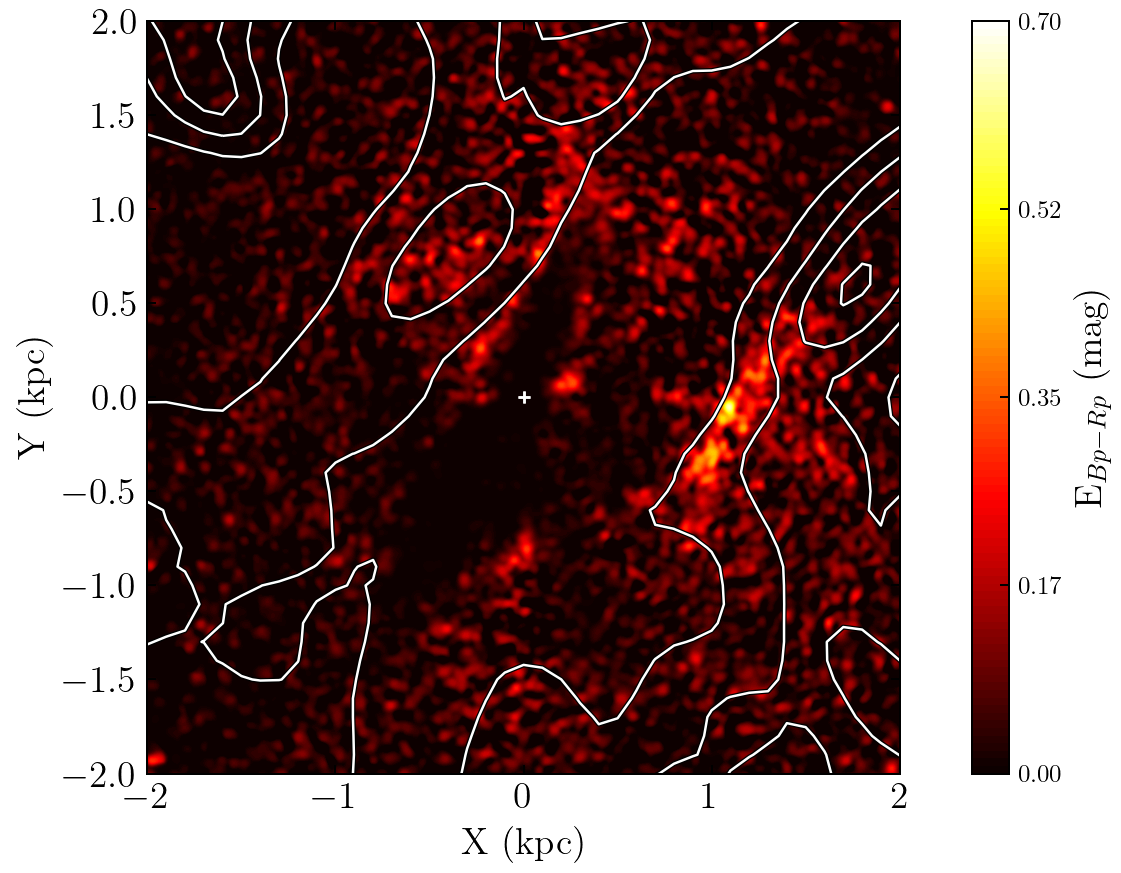} 
\end{subfigure}
\begin{subfigure}{0.5\textwidth}
\includegraphics[width=0.95\linewidth]{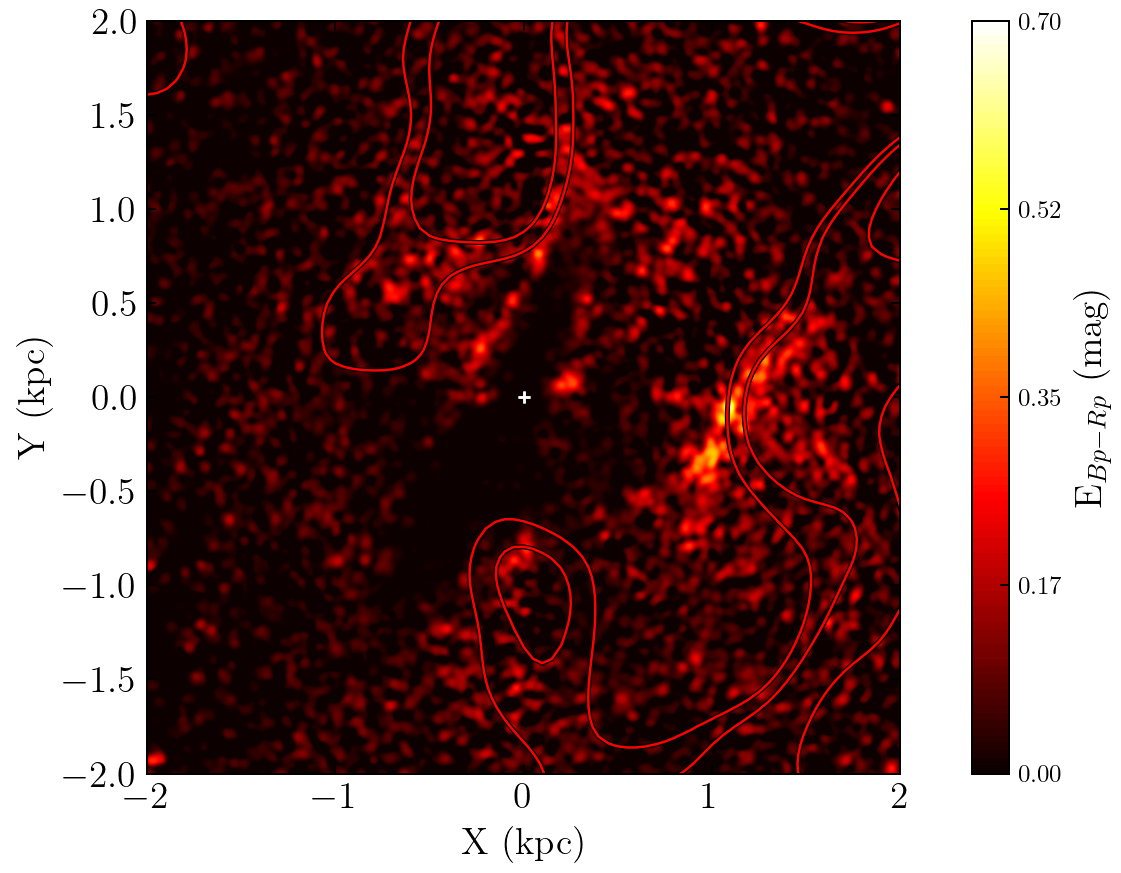}
\end{subfigure}
\begin{subfigure}{0.5\textwidth}
\includegraphics[width=0.95\linewidth]{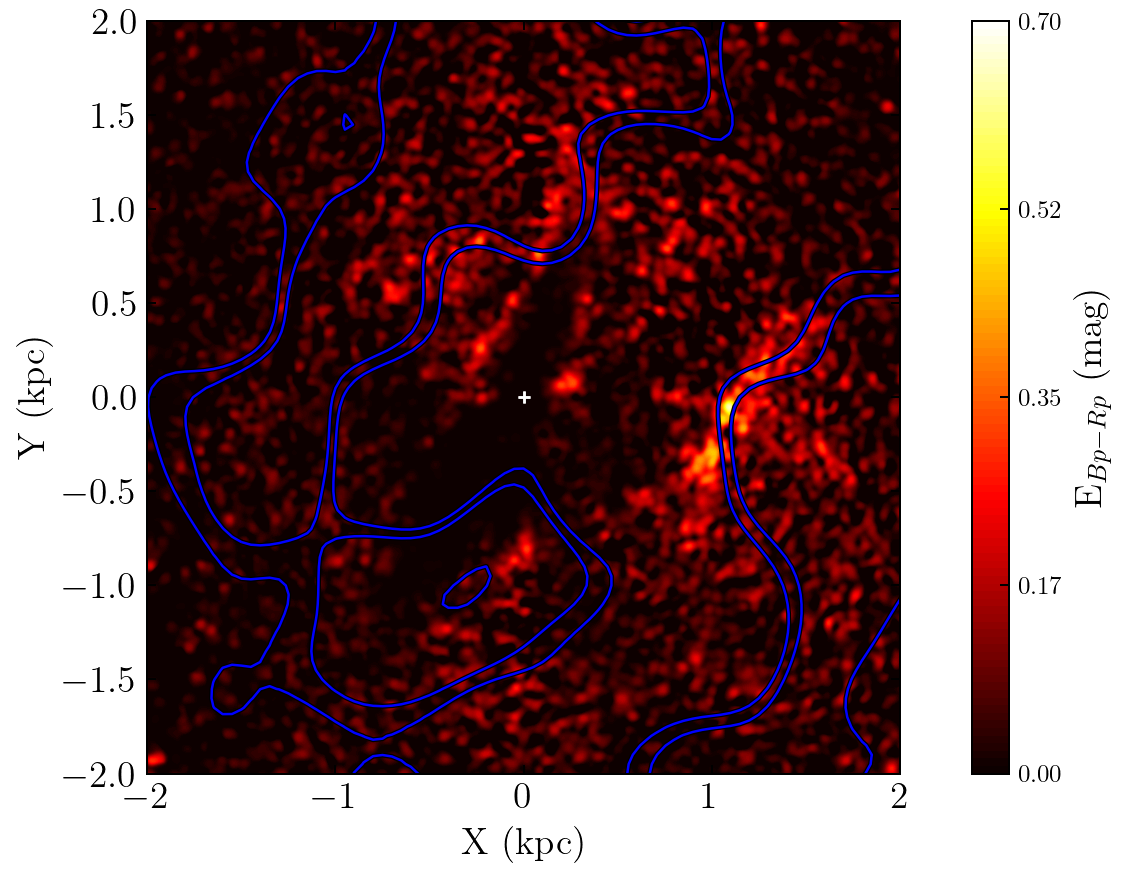}
\end{subfigure}
    \caption{2D $XY$ projections of the extinction overlaid with different type of contours. The Sun is at coordinates (X, Y) = (0, 0) represented by a white cross and the Galactic centre direction is to the right. The upper left panel shows $\Sigma$H$_{tot}$ surface density contours from \cite{Mertsch_2023}, the upper right panel shows the upper main sequence (UMS) stars density contours from \cite{spiral_arms_EDR3}, the bottom left panel shows the enhancement of \mh\ from \cite{Barbillon24} and the bottom right panel shows the deficiency of \cafe\ from \cite{Barbillon24}.}
\label{3D_Extinction_maps_comparison_contours}
\end{figure*}

Moreover, we overplotted star-forming regions from \cite{Zucker_2021, Zucker_2022}\footnote{See their intuitive \href{https://faun.rc.fas.harvard.edu/czucker/Paper_Figures/Interactive_Figure1.html}{3D interactive figure}. They reported a detailed analysis of the 3D positions, shapes, and motions of dense gas and young stars within 200 pc of the Sun.} which match fairly well the distribution of the dust. The Split structure (cf. the blue line in the right panel of Fig. \ref{3Dmaps_KDE_sunvicinity_XY}) is composed of the molecular clouds of Chamaeleon, Musca, Lupus, Pipe, Corona Australis, and Ophiuchus. In Fig. \ref{LB_3D_map}, these molecular clouds traced by blue and green markers, show an arm-like feature that bridges the Local and Sagittarius-Carina arms. It is well identified as a linear elongated molecular complex. On the other hand, the RW (cf. the magenta line in the right panel of Fig. \ref{3Dmaps_KDE_sunvicinity_XY}) is composed of the Cepheus, Perseus, Taurus and Orion molecular clouds and are referred as red, yellow and pink markers in Fig. \ref{LB_3D_map}. The structure is vertically undulated, putting in evidence the wavy distribution of the dust. This pattern is also found in the 3D density distribution of young stars from \cite{Zari18} based on the \Gaia~DR2 data. They showed that the wavy pattern is dominated by young star clusters in dense and compact clumps, but also surrounded by older sources, whose distribution is more diffuse. Our maps agree well with \cite{Zari18} maps, except for the Vela cloud location. This difference likely stems from the use of \Gaia\ DR2 distances in \cite{Zari18}, compared to the EDR3-based distances from \cite{Bailer-Jones_distances} adopted here. \cite{DR3Gaia_2023} summarised the improvement between the different data release.

Finally, our projections at small and larger scales, show that the RW and Split features are composed of discontinuous structures in agreement with previous works such as \cite{Zari18, Alves_2020, Zucker_2022}.  This calls into question the continuous vision of Gould belt structure.

\section{Discussion and conclusions}
\label{Discussion}

We presented new 2D and 3D highly resolved extinction maps derived from the last \Gaia\ DR3 \gspspec\ spectroscopic survey. The extinction values are only based on homogeneous data from the \Gaia\ space mission; more precisely, on stellar atmospheric parameters as they are not affected by extinction. Moreover, unlike other studies in the literature, we avoid the use of priors or likelihood analyses for our reddening estimation. Although the discretisation method used to compute the differential extinction $\Delta E$, and the fixed resolution can be discussed and improved, the general agreement between this work and the literature is proving the reliability of the derived \ebprp\ extinctions and the added value of this complementary approach. Our all-sky maps (cf. Figs. \ref{lvsb_plot} and \ref{lvsb_plot_d_bins}) illustrate the high quality flags sample, while the 2D projections (cf. Figs \ref{3Dmaps_KDE_sunvicinity_XY} to \ref{LB_3D_map}) highlight a MW disc sample, including high quality astrometric parameters and flags. We have created new 3D dust maps at different spatial scales and resolutions around the Sun. The larger scale covers a volume of 4$\times$4$\times$0.8 kpc in Galactic Cartesian coordinates with a discretisation applied in the spherical coordinates of ($dr, d\theta, d\phi) = (40~\text{pc}, 1\deg, 1\deg)$ (cf. Sect. \ref{3Dmaps_section_sunvicinity}). In addition, the smaller scale covers a volume of 1$\times$1$\times$0.8 kpc in Galactic Cartesian coordinates with a discretisation applied in the spherical coordinates of ($dr, d\theta, d\phi) = (30~\text{pc}, 1\deg, 1\deg)$ (cf. Sect. \ref{LB_section}). The 3D extinction datasets for the different spatial coverage around the Sun are available to download via Zenodo at \url{https://doi.org/10.5281/zenodo.17598127}.

Our maps provide a crucial and new understanding of the structure of the MW ISM and the processes that shaped it. Figure \ref{3D_Extinction_maps_comparison_contours} shows an overview of the dust distribution on the \textit{large scale} compared to different spiral arm tracers. First, the gas density observations of HI, H$_2$, and ionised hydrogen (HII) are the most important tracer of the structure and dynamics of the ISM \citep{Kewley2019, Mertsch_2023}. Spiral arms, being gas-rich regions and major sites of star formation, are naturally expected to exhibit a gas distribution that closely traces the dust. Recently, \cite{Soding_2025} published updated 3D distributions of neutral atomic hydrogen (HI) and molecular hydrogen\footnote{Estimated from the linear relation between the H$_2$ column density NH$_2$ and the velocity-integrated CO brightness temperature.} (H$_2$) using a Bayesian reconstruction of Galactic gas densities from the HI4PI survey \citep{HI4PI}. 
The upper left panel  of Fig.~\ref{3D_Extinction_maps_comparison_contours} presents a comparison between our extinction map and the gas distribution of \cite{Soding_2025}. In particular, we have used the $\Sigma$H$_{tot}$ surface density estimated from the total hydrogen density $n^{tot} = n^{HI} + 2\times n^{H_2}$\footnote{\url{https://zenodo.org/records/12578443}.}. The two patterns are in good overall agreement, similar to the extinction distribution shown in Fig.~21 of \cite{Soding_2025} based on \cite{Edenhofer_2024}.

Secondly, it is interesting to compare our extinction distribution to the overdensity contours of young upper main sequence (UMS) stars \citep{spiral_arms_EDR3}, as shown in the upper right panel of Fig.~\ref{3D_Extinction_maps_comparison_contours}. Several studies have shown that star formation tends to occur preferentially along the leading edges of spiral arms in external galaxies \citep{Seigar2002, Martinez_2009, Silva_villa_2012}. This phenomenon is consistent with the extinction distribution in the Sagittarius-Carina arm, where the highest values are found along the leading edge. 
In contrast, the Local Arm presents higher extinction patterns along the trailing edge. Interestingly, using Gaia DR3 data of open clusters, \cite{CastroGinard21} found that the speed of the Local Arm pattern exceeds the MW rotation curve, while that of Sagittarius lags behind.  In addition, the Local arm presents a sharp quasi-parallel offset with the RW and the \cite{spiral_arms_EDR3} overdensity contours, as already noticed by previous works \citep{Alves_2020, Swiggum2022}. In this sense, \cite{Martinez-medina2025} suggests that the RW is the gas reservoir of the Local arm. If confirmed, this would be consistent with the density wave model of the spiral arms \citep{Lin1964, Toomre1969} which suggests that the gas experiences compression when it enters the density wave, leading to enhanced star formation on the edge of the pattern.
However, the dust distribution in the proximity of the Sun could have different origins \citep[see for instance][who summarised different formation scenarios of the Gould Belt]{Zari18}. In particular, as in the case of the spiral arms formation processes, perturbations induced by external galaxies have to be taken into account. \cite{Bland-hawthorn19} suggests to consider the 3D dust distribution in the larger framework of disc corrugations and phase-space perturbations. ISM corrugations have been observed in nearby galaxies experiencing merging processes 
\citep{Edelshon_1997, Matthews_2008}. In this sense, \cite{Thulasidharan2022} used N-body simulations to interpret the kinematics of the RW, suggesting that it could be the response of the disc to an external perturbation similar to the Sagittarius dwarf galaxy. 
Moreover, \cite{Alves_2020} proposes that a tidally stretched gas cloud accreted by the Galactic disc, could reproduce the observed shape and damped oscillation. However, this scenario would require the clouds to be in synchrony with the Galactic rotation (including a cloud's $V_{LSR} \sim$ 0 km/s).

Finally, the bottom panels of Fig. \ref{3D_Extinction_maps_comparison_contours} present the comparison between the extinction distribution and two chemical patterns of the spiral arms: the metallicity enhancement traced by the Gaia  \mh$_{\gspspec}$ parameter \citep{spiral_arms_DR3, Barbillon24} (bottom left panel), and the deficiency in $\alpha$-elements with respect to iron, traced by the Gaia \cafe$_{\gspspec}$ abundances \citep{Barbillon24} (bottom right panel). It is worth noting that several regions of the map are consistently identified across the different tracers. 
In particular, the features around the Sun, the LB, the RW, the Split, the Cepheus spur and Sagittarius-Carina spur match perfectly with the $\Sigma$H$_{tot}$ surface density maps, as expected. These regions are also found in the density distribution of UMS stars, the \mh\ enrichment and \cafe\ deficiency of the spiral arms, unveiling the link between the distribution of the dust, the gas, the position of stars and their chemical evolution. 

From a more global perspective, the densest dust structures located in the Vela cloud around ($X$, $Y$) $\sim$ (-0.05, -0.95) kpc, the Sagittarius-Carina arm and part of the Local arm match the different referenced contours. Conversely, thanks to the different tracers, a large under-density is observed around ($X$, $Y$) $\sim$ (1, -1) kpc, which has also been highlighted by \cite{Dharmawardena24} and citations therein. In addition, we note that the extinction tends to be higher at positive azimuthal angles $\phi$. These azimuthal variations could be a consequence of the spiral arms, since similar variations are also visible. Indeed, these fluctuations are also observable in the gas density maps of HI, H$_2$ and HII \citep{Mertsch_2023, Rezaei_24, Soding_2025}, 
in the Gaia chemical pattern of the spiral arms \citep{spiral_arms_DR3,Barbillon24}, the kinematics \citep{Gaia_Drimmel_2023}, the dynamics \citep{Palicio_2023} of stars, the variation of the star formation rate over the MW \citep{Elia22}, and the dynamical estimates of the disc surface density \citep{Widmark24}. This azimuthal dependence, together with the connection between dust, gas, stars, and their chemical composition, reveals the physical mechanisms driving the formation of the spiral arms, thus highlighting the complex Galactic ecology. Hence, these data provide crucial constraints to Galactic simulations including both gas and stars \citep[see for instance][]{Nexus, Newhorizon} and will be further investigated under this perspective in future work. 



\begin{acknowledgements}
    This work presents results from the European Space Agency (ESA) space mission \textit{\Gaia}\ (\url{https://www.cosmos.esa.int/gaia}). \textit{\Gaia}\ data are processed by the \textit{Gaia} DPAC. Funding for the DPAC is provided by national institutions, in particular the institutions participating in the \textit{\Gaia}\ MultiLateral Agreement (MLA). The \textit{\Gaia}\ archive website is \url{(https://archives.esac.esa.int/gaia)}. MB and PAP acknowledge financial supports from the French Space Agency, Centre National d’Études Spatiales (CNES). ARB and PdL acknowledge funding from the European Union’s Horizon 2020 research and innovation program under SPACE-H2020 grant agreement number 101004214 (EXPLORE project). We would like to thank the Conseil régional Provence-Alpes-Côte d'Azur for its financial support.
\end{acknowledgements}

\bibliographystyle{aa}  
\bibliography{main} 

\appendix

\section{Stellar density distribution }
\label{AppendixA}

\begin{figure*}[htbp]
\includegraphics[width=1\linewidth]{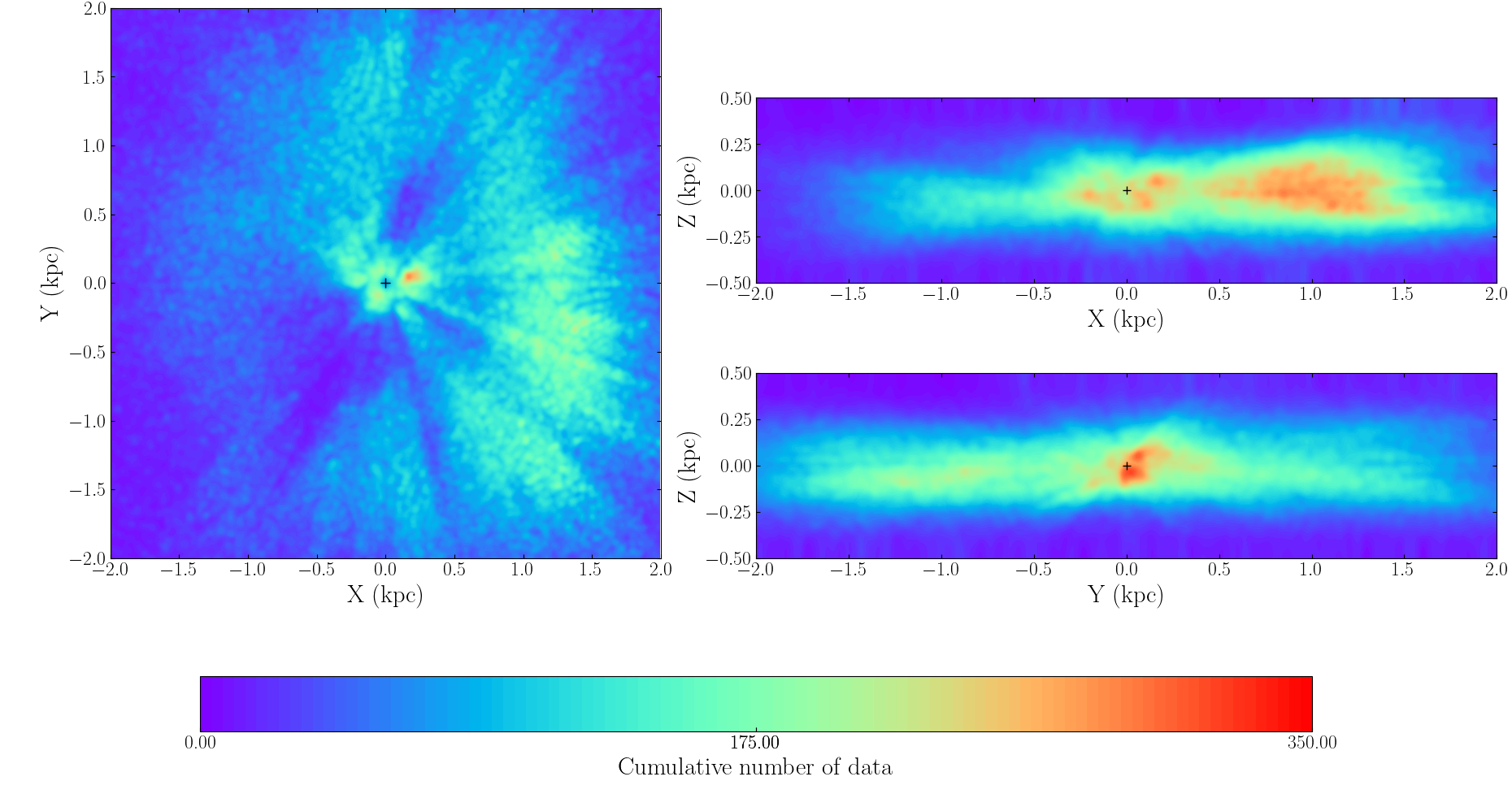}
\caption{Stellar density distribution in the 2D ($X$, $Y$, $Z$) projections of the 3D map (cf. section \ref{3Dmaps_section}). The Sun is at coordinates ($X$, $Y$, $Z$) = (0, 0, 0.021) kpc represented by a black cross and the Galactic centre direction is to the right.}
\label{stat_3Dmaps_KDE_sunvicinity}
\end{figure*}

Figure \ref{stat_3Dmaps_KDE_sunvicinity} presents the 2D projection of the distribution of the number of stars in the Galactic disc by accumulating the number of stars through 0.8 kpc in the $Z$ direction (left panel), 4 kpc in the $Y$ direction (middle panel) and 4 kpc in the $X$ direction (right panel). We have set the $h$ bandwidth to $h$=20 pc.

\end{document}